\documentclass[onecolumn, superscriptaddress, secnumarabic,amssymb, nobibnotes, aps, prl]{revtex4-2}
\usepackage{epsf}
\usepackage{graphicx}
\usepackage{sidecap}
\usepackage{amsmath}
\usepackage{float}
\usepackage{hyperref}
\hypersetup{
	colorlinks=true,
	citecolor=red,
	linkcolor=blue,
	filecolor=blue,   
	urlcolor=blue,
}
\renewcommand{\thefigure}{{\textbf{\arabic{figure}}}}

\makeatletter
\def\maketitle{
\@author@finish
\title@column\titleblock@produce
\suppressfloats[t]}
\makeatother

\begin{document}

\title{Temperature Dependent Evolution of the Electronic Structure in EuZn$_2$As$_2$ across the N\'eel Transition}%

\author{Milo~Sprague}
\affiliation{Department of Physics, University of Central Florida, Orlando, Florida 32816, USA}

\author{Anup~Pradhan~Sakhya}
\affiliation{Department of Physics, University of Central Florida, Orlando, Florida 32816, USA}
\affiliation{Research Institute for Synchrotron Radiation Science, Hiroshima University, Higashi-Hiroshima 739-0046, Japan}

\author{Barun~Ghosh}
\affiliation{Department of Physics, Northeastern University, Boston, Massachusetts 02115, USA}
\affiliation{Quantum Materials and Sensing Institute, Northeastern University, Burlington, Massachusetts 01803, USA}
\affiliation{Department of Condensed Matter and Materials Physics, S.N. Bose National Center for Basic Sciences, Kolkata 700106, India}

\author{Mazharul~Islam~Mondal}
\affiliation{Department of Physics, University of Central Florida, Orlando, Florida 32816, USA}

\author{Arun~K.~Kumay}
\affiliation{Department of Physics, University of Central Florida, Orlando, Florida 32816, USA}

\author{Himanshu~Sheokand}
\affiliation{Department of Physics, University of Central Florida, Orlando, Florida 32816, USA}

\author{Kapil Gope}
\affiliation{Department of Condensed Matter and Materials Physics, S.N. Bose National Center for Basic Sciences, Kolkata 700106, India}

\author{Tetiana~Romanova}
\affiliation{Institute of Low Temperature and Structure Research, Polish Academy of Sciences, ul. Ok\'olna 2, PL-50-422 Wroc\l aw, Poland}

\author{Dariusz~Kaczorowski}
\affiliation{Institute of Low Temperature and Structure Research, Polish Academy of Sciences, ul. Ok\'olna 2, PL-50-422 Wroc\l aw, Poland}

\author{Arun~Bansil}
\affiliation{Department of Physics, Northeastern University, Boston, Massachusetts 02115, USA}
\affiliation{Quantum Materials and Sensing Institute, Northeastern University, Burlington, Massachusetts 01803, USA}

\author{Madhab~Neupane}\email{Corresponding author: madhab.neupane@ucf.edu}
\affiliation{Department of Physics, University of Central Florida, Orlando, Florida 32816, USA}

\begin{abstract}
	
	Magnetoresistive materials have been tremendously important for the development of magnetic memory storage and spintronic devices. Recently, the antiferromagnetic Eu\textit{X$_2$Pn$_2$} compounds, with X being a transition metal and Pn being a pnictogen, have seen intensive research interest due to their unusual anomalous Hall effect behavior and pronounced resistive anomaly near the N\'eel temperature (T$_\text{N}$). These magnetotransport phenomena have been interpreted in the context of short-ranged ferromagnetic fluctuations, magnetic polaron formation, canted spin configurations, and temperature-dependent metal-insulator transitions in the electronic structure. Here, we report the observation of such a pronounced resistivity anomaly in EuZn$_2$As$_2$ near T$_\text{N}$ = 19 K. We demonstrate the suppression of this anomaly using applied magnetic fields, both in-plane and out-of-plane. To further interpret the origin of the observed transport behavior, we studied the temperature-dependent electronic structure using combined angle-resolved photoemission spectroscopy (ARPES) and first-principles density functional theory (DFT) calculations, which exhibits limited modifications to the bands across T$_\text{N}$ away from the Fermi energy. This lack of involvement of the electronic structure indicates a spin-scattering origin of the aforementioned transport properties, rather than a reconstruction of the Fermi surface.
	
\end{abstract}
\maketitle

\section{Introduction}

The interplay of magnetism and electronic transport of materials has been a highly pursued research initiative in recent years. The development of spintronics and magnetic storage devices has increasingly highlighted magnetoresistance (MR) for magnetic reading and controlling spin currents \cite{RamirezCMR,FertSpintronics}. The family of Eu-ternary pnictides, having stoichiometry Eu$X_2Pn_2$ (X = transition metal and Pn = pnictogen), has been discovered to host negative MR \cite{Blawat,TaftiAnisotropy,LuoMREuZn2As2,RahnCouplingEuCd2As2,DuConsecutiveEuCd2As2,WangMREuCd2As2,ZhangMIT,RahmanEuCd2P2MR}, anomalous Hall and anomalous Nernst effects \cite{RahnCouplingEuCd2As2,XuEuCd2As2Hall,YiEuZn2As2Hall,RoychowdhuryAnomalousEuCd2As2} in both antiferromagnetic (AFM) and paramagnetic (PM) phases, magnetic polaron formation \cite{KrebberEuCd2P2MagPol, ZhangMagPolaron,RosaEu5In2Sb6Polaron}, canted spin configurations \cite{BukowskiAFMOrder,TaddeiSpinCantedWeyl}, and fluctuating ferromagnetic order in the PM phase \cite{Goryunov2012esr, Goryunov2014spin, MaEuCd2As2Fluctuation,SpragueFluctuationsEuZn2Sb2}.

\indent The influence of low-temperature AFM ordering upon the electronic band topology and anomalous carrier transport has been the center of discussions in these systems. The most prominent example highlighting the intrigue of this material family is EuCd$_2$As$_2$, which was initially brought to wider attention as an example AFM Dirac semimetal candidate \cite{RahnCouplingEuCd2As2,Ma2020EuCd2As2AFM}. Previous theoretical calculations on EuCd$_2$As$_2$ have shown that the material transitions from a topological insulator to a Dirac semimetal between in-plane and out-of-plane AFM moment configurations \cite{WangEuCd2As2CalcWeyl}. Transitioning to ferromagnetic ordering causes the splitting of these Dirac points, transforming them into a single pair of Weyl points \cite{WangEuCd2As2CalcWeyl}.\\
\indent EuZn$_2$As$_2$ is a notable AFM semiconductor within this family which shows negative MR, anomalous/topological Hall effects, and fluctuating short-ranged magnetic correlations \cite{LuoMREuZn2As2, YiEuZn2As2Hall, BukowskiAFMOrder,Goryunov2012esr, Goryunov2014spin}. Scattering of itinerant electrons from these short-ranged magnetic interactions \cite{YamadaAFMMagnetoresistance, UsamiAFMMR} have been widely considered as responsible for the MR and Hall behavior in this system \cite{Blawat,TaftiAnisotropy,LuoMREuZn2As2, YiEuZn2As2Hall}. However, the possibility of temperature-induced electronic structure changes around the AFM transition introduces a potentially complex interdependence between electronic and magnetic properties, as has been implicated in EuCd$_2$P$_2$ and EuZn$_2$P$_2$ \cite{SinghSuperexchange, ZhangMIT,EuZn2P2FMMetal}. Here, we elucidate the magnetic and electronic interplay in EuZn$_2$As$_2$ by examining the magnetoresistive properties of this system and by investigating the electronic structure across the PM-to-AFM transition at the N\'eel temperature (T$_N$ = 19 K) using combined angle-resolved photoemission spectroscopy (ARPES) measurements and first-principles density functional theory (DFT) calculations. We find a significant resistive anomaly at T$_\text{N}$ where magnetic fluctuations appear to increase the resistivity substantially. Application of external magnetic fields along either the in-plane or out-of-plane axes results in a suppression of this resistive anomaly. Our ARPES-measured band structure reveals a minor shifting of the valence band energies in response to the onset of magnetic ordering. However, these changes do not appear to extend to the Fermi energy, indicating that a significant reconstruction of the Fermi surface is not present.\\
\section{Methods}
High-quality EuZn$_2$As$_2$ single crystals were grown by using Zn-As flux and characterized using X-ray diffraction and energy-dispersive X-ray spectroscopy. Magnetic properties were investigated in the temperature range 1.72-400 K and in magnetic fields up to 9 T using a Quantum Design MPMS-XL superconducting quantum interference device (SQUID) magnetometer. Electrical transport studies were performed from 2 to 300 K in magnetic fields up to 9 T using a Quantum Design PPMS-9 platform. The electrical leads were made of gold wires attached to the bar-shaped specimens with silver-epoxy paste. The experiments were done employing a standard four-point ac technique. \\
\indent Density functional theory (DFT) based calculations were performed using projector-augmented wave (PAW) pseudopotentials implemented within the Vienna Ab initio Simulation Package (VASP) \cite{ModifiedPAW,IterativeVASP,VASPMolDyn}. Band structure calculations were performed using the Perdew-Burke-Ernzerhof (PBE) functional \cite{HSE} with a plane-wave cutoff of 400 eV. A $\Gamma$-centered 12$\times$12$\times$6 k-mesh was used for the Brillouin zone integration. The cell parameters and the internal atomic coordinates were optimized fully until the force acting on any atom became less than 0.001 eV/$\text{\AA}$. The collinear magnetic calculations were performed using appropriate pseudopotentials within the DFT+U framework, with an effective on-site Coulomb interaction $U_{\mathrm{eff}} = 5$ eV, following the Dudarev approach as implemented in VASP \cite{DFTU}. ARPES measurements were performed at the Advanced Light Source Beamline 4.0.3. Sample temperatures of 13.5 K and 30 K were used for comparison of the AFM and PM electronic structures, respectively. During these measurements the pressure in the main chamber was maintained to better than $1\times10^{-10}$ Torr. The angular and energy resolutions were set to better than $0.2^{\circ}$ and 15 meV, respectively. Photon energies between 40 eV and 100 eV were employed during the ARPES measurements.\\


\begin{figure*}
	\centering
	\includegraphics[width=1\linewidth]{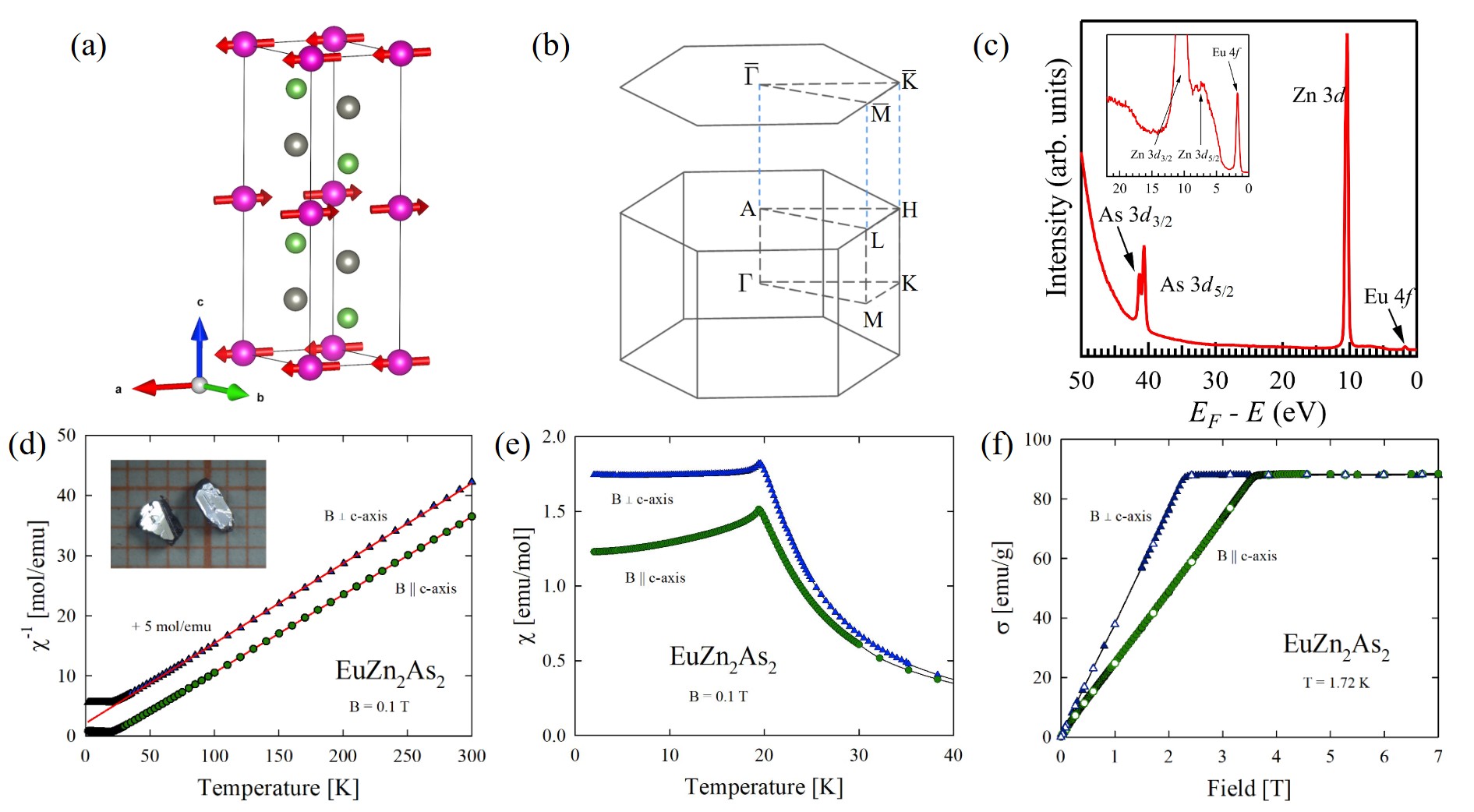}
	\caption{Crystal structure, core-level photoemission, and magnetic characterization of EuZn$_2$As$_2$. (a) Trigonal crystal structure consisting of Eu, Zn, and As atoms represented in purple, silver, and green, respectively. The in-plane orientation of the Eu magnetic moments employed in our AFM calculations are indicated by purple arrows. (b) Corresponding Brillouin zone (BZ) with the high-symmetry points indicated. The lower BZ depicts the 3D bulk shape, whereas the surface (001) BZ is projected above. (c) The core-level photoemission reveals the presence of Eu 4$f$, Zn 3$d$, and As 3$d$ peaks, indicated with arrows. The inset panel on the top-left is zoomed into lower binding energies to highlight the spin-orbit split Zn 3$d$ levels and Eu 4$f$ peak positions. (d) Inverse-magnetic susceptibility of EuZn$_2$As$_2$ for magnetic fields $B_{\perp}$ (blue triangle points) and $B_{\parallel}$ (green circular data points). Curie-Weiss fits are performed for the $B_{\perp}$ and $B_{\parallel}$ field orientations and are plotted in red. Inverse magnetic susceptibility measurements were performed under a field $B=0.1~\text{T}$. The inset shows the high-quality single crystals of EuZn$_2$As$_2$ (e) Temperature dependence of the magnetic susceptibility for $B_{\perp}$ (blue) and $B_{\parallel}$ (green) measured under an applied field of $B=0.1~\text{T}$. (f) Magnetization vs applied field for both $B_{\perp}$ (blue) and $B_{\parallel}$ (green) measured at a temperature of 1.72 K.}
	\label{fig:ezafigure1v1}
\end{figure*}
\section{Results}
EuZn$_2$As$_2$ crystallizes in the trigonal space group \textit{P$\overline{3}$m1} (No. 164) with Eu atoms defining the corners of the unit cell (Figure 1(a)). Sandwiched between the trigonal Eu layers are alternating zigzagged ZnAs layers arranged in a honeycomb pattern. The trigonal lattice produces the hexagonal Brillouin zone (BZ), shown in Figure 1(b), with high symmetry points indicated. The BZ surface-projection along the cleaved (001) direction is shown above the bulk one. In Figure 1(c), we present the core-level photoemission spectrum for EuZn$_2$As$_2$ which shows prominent Eu 4$f$, Zn 3$d$, and As 3$d$ peaks. The presence of the Eu 4$f$ peak (see the inset of Figure 1(c)) at a binding energy of about $E-E_F = 1.7~\text{eV}$, highlights the divalent occupancy of the Eu 4$f$ band \cite{Eu4fValency}.

To establish the magnetic behavior of EuZn$_2$As$_2$, we have performed temperature-dependent magnetic susceptibility (Figures 1(d,e)) and field-dependent magnetization measurements (Figure 1(f)) for applied fields both perpendicular (B$\perp$c) and parallel (B$||$c) to the c-axis. Curie-Weiss behavior is observed within the PM inverse-susceptibility character presented in Figure 1(d) above T = 40 K. A linear fit is performed on the inverse susceptibility shown in Figure 1(d), where from the slope we have extracted an effective magnetic moment of $\mu_{eff}^{||}$ = 7.86 $\mu_B$ and $\mu_{eff}^{\perp}$ = 7.76 $\mu_B$ \cite{Blawat, TaftiAnisotropy, BukowskiAFMOrder, LuoMREuZn2As2, EuZn2As2Optical}. These moments are close to the theoretical value of 7.94 $\mu_B$ \cite{CurieTutorial}, in agreement with the divalent Eu state corresponding to $g=2$ and $J=7/2$, further corroborating our core-level photoemission results (Figure 1(c)). From the temperature-axis intercept, we have obtained a positive effective PM Curie temperature of $\theta_p^{||}$ = 18.7 K and $\theta_p^{\perp}$ = 21.7 K. Similar positive-valued PM Curie temperatures from both field orientations indicate the presence of an isotropic ferromagnetic exchange interaction between Eu-moments within the PM phase \cite{Blawat, CurieTutorial}. Despite the apparent FM-like exchange in the PM phase indicated by the effective Curie temperatures, the low temperature magnetic susceptibility measurements show a suppression below T$_N$ = 19.4 K, consistent with AFM ordering for low temperatures. Figure 1(e) presents the magnetic susceptibility as a function of temperature from 2 K to 40 K in the presence of a magnetic field $B=0.1~\text{T}$. We observe increasing susceptibility with decreasing temperature, with a pronounced peak at T$_\text{N}$ = 19.4 K for both $B_{\perp c}$ and $B_{\parallel c}$. Below this peak, the susceptibilities show reduced temperature dependence with an observable anisotropy; however, it's been noted that the anisotropy in the magnetic response is considerably weaker than in the EuCd$_2$\textit{Pn}$_2$ materials, likely due to weakened spin-orbit interactions and greater localization of the $d$-orbitals in EuZn$_2$As$_2$ \cite{TaftiAnisotropy}. Figure 1(f) presents the magnetization as a function of applied magnetic field for a low-temperature isotherm of T = 1.72 K. For both $B_{\perp c}$ and $B_{\parallel c}$ we find increasing magnetization over weak applied fields. The $B_{\perp c}$ magnetization is found to saturate at a critical field of 2.3 T, where the $B_{\parallel c}$ magnetization saturates at a critical field of 3.6 T. Considering the stronger susceptibility within the $\perp c$ relative to the $||c$ direction, the positive isotropic effective Curie temperatures, and the Eu-moment response consistent with an Eu$^{2+}$ valency, the magnetism in EuZn$_2$As$_2$ is best described as a type-A AFM with the Eu$^{2+}$ moments oriented predominantly within the $ab$-plane for temperatures below T$_N$ = 19.4 K. This magnetic configuration is visualized by the moments depicted with purple arrows in Figure 1(a). Above T$_N$, ferromagnetic exchange interactions dominate, potentially through persistent magnetic fluctuations.\\
\indent The presence of the AFM transition at T$_\text{N}$ = 19.4 K has a profound effect on the temperature-dependence of the longitudinal resistivity, as shown in Figure 2(a,b). Above 100 K, the  fairly high resistivity increases with increasing temperature, which establishes the semimetallic behavior of our EuZn$_2$As$_2$ sample. This indicates a small number of holes were introduced during the synthesis process, as non-magnetic DFT calculations predict a semiconducting gap at the chemical potential \cite{TaftiAnisotropy, EuZn2As2Optical}. A sharp resistivity peak is observed at T$_\text{N}$, which shows an extended tail in the PM phase up to 100 K. To assess the origin of this large resistivity at the transition temperature, we repeated resistivity measurements for various applied magnetic fields, as indicated in Figure 2(a,b). The application of both $B_{\parallel}$ (Figure 2(a)) and  $B_{\perp}$ (Figure 2(b)) results in the suppression of the resistivity peak, which is more pronounced for magnetic fields applied  $B_{\parallel}$ to \textit{c} when compared with $B_{\perp}$ to the \textit{c}-axis. Further, we can observe the resistivity peak shift toward lower temperatures with the increase in magnetic field for both $B_{\parallel}$ and  $B_{\perp}$, indicating the common suppression of the magnetic transition by applied fields. However, we also find an extra bump that emerges in the PM state, which may be associated with spin fluctuations \cite{Blawat, TaftiAnisotropy}.

A more detailed description of the magnetoresistive behavior is obtained by plotting MR isotherms against an applied magnetic field. For small fields ($B_{\parallel}$ and $B_{\perp}$ $<$ 0.5 T), a positive MR of the in-plane current, $I_{\perp c}$, is observed within the AFM phase (Figures 2(c,d)). Upon increasing the field strength further, the low-temperature (2 K) resistivity isotherm decreases continuously up until the critical magnetization ($B_{\parallel}$ = 3.6 T and $B_{\perp}$ = 2.3 T) is reached, where a sudden cusp is observed, coinciding with the magnetic saturation. Increasing the temperature for both $B_{\parallel}$ and $B_{\perp}$ reduces the critical field as indicated by the shifting of this cusp toward lower fields.  Beyond the critical field application, in which the magnetic moments are saturated, the resistivity shows a weakly increasing MR. Qualitatively, this behavior persists upon increasing the temperature of the magnetically ordered sample, however we find (a) a reduction in the critical field value, (b) a smoothing of the MR discontinuity at said critical field, and (c) an increasingly negative MR upon approaching the AFM phase transition temperature \cite{YiEuZn2As2Hall}. This negative MR persists into the PM phase for both $B_{\parallel}$ and $B_{\perp}$ field orientations, as demonstrated in Figures 2(e,f). This negative MR is reduced as temperature is raised further past T$_N$ \cite{TaftiAnisotropy, YiEuZn2As2Hall}. 

\begin{figure*}
	\centering
	\includegraphics[width=1\linewidth]{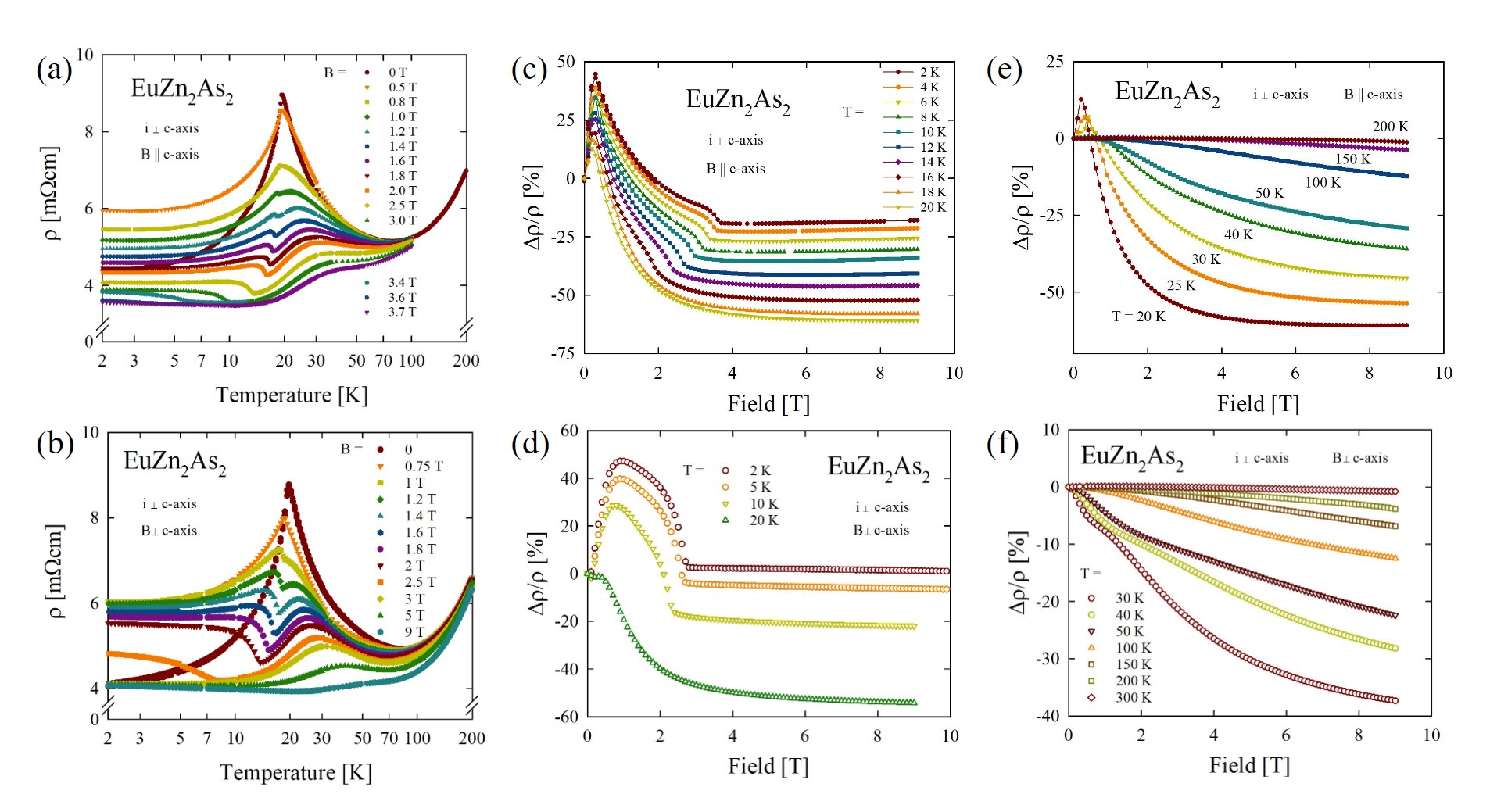}
	\caption{In-plane current MR behavior of EuZn$_2$As$_2$ single crystals. (a,b) Temperature variation of longitudinal resistivity for B$||$c and B$\perp$c, respectively. The B$||$c fields range in strength from 0 to 3.7 T, while B$\perp$c ranged from 0 to 9 T. (c,d) B$||$c and B$\perp$c MR isotherms within the AFM phase, respectively, plotted for temperatures between 2 K and 20 K. (e.f) MR isotherms within the PM phase for temperatures ranging between 20 K and 200 K for B$||$c and B$\perp$c applied fields, respectively.}
	\label{fig:ezafigure2v1}
\end{figure*}

\begin{figure*}
	\centering
	\includegraphics[width=1\linewidth]{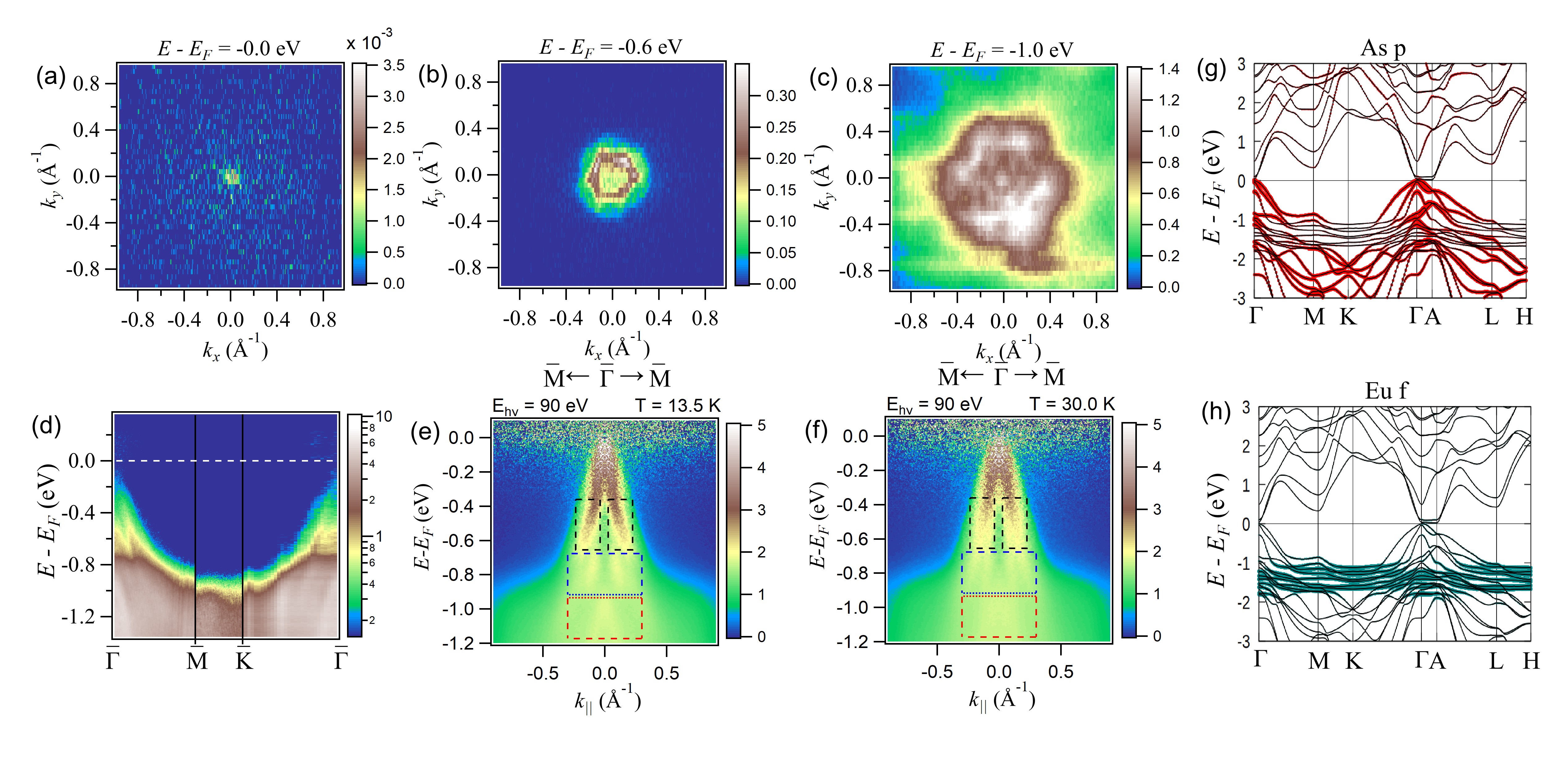}
	\caption{Fermi surface and CECs in EuZn$_2$As$_2$. (a-c) Constant energy contours (CECs) taken at binding energies between $E-E_F=-0.2~\text{eV}$ and $E-E_F=-1.0~\text{eV}$. (d) Electronic band dispersion along the three surface high-symmetry directions. (e,f) ARPES-obtained dispersion along the $\overline{\text{M}}-\overline{\Gamma}-\overline{\text{M}}$ direction at temperatures of 13.5 K (AFM) and 30 K (PM), respectively. (g,h) AFM-phase band dispersion along the bulk high-symmetry directions. Orbital projections onto the As $p$ (e) and Eu $f$ (f) bands are shown in red and teal, respectively. The ARPES results presented in panels (a-d) were measured using 100 eV photon energy, while those in (e,f) were measured using 90 eV photons.}
	\label{fig:ezafigure3v4}
\end{figure*}

To compliment the unusual resistivity and magnetoresistence results, we have performed high-resolution ARPES measurements measured on the cleaved (001) surface and DFT-based first-principles calculations to study the electronic band structure of EuZn$_2$As$_2$. The nature of this (001) surface has been described by recent STM/STS measurements, which demonstrate that cleavage produces clean Eu or ZnAs layers \cite{EuZn2As2_Large_Bandgap}. We begin with the Fermi surface and constant energy contours (CECs) of EuZn$_2$As$_2$ obtained through ARPES measurements using 100 eV incident photon energy, measured at a temperature of 13.5 K (Figure 3(a-c)). The Fermi surface (Figure 3(a)) forms a nearly vanishing circular pocket surrounding the $\overline{\Gamma}$ point. This small Fermi surface is likely generated from a slight hole-doping of our sample during synthesis, as indicated by the resistivity measurements; however, recent experimental works also demonstrate the vanishing of the semiconducting gap across the N\'eel transition \cite{EuZn2As2Optical, EuZn2As2_Large_Bandgap, Quantum_Limit}. This pocket expands in area over increasing binding energy, demonstrating the hole-like nature of these bands. By the $E-E_F\approx-0.6~\text{eV}$ CEC, we can resolve a hexagonal reshaping of the pocket as binding energy is increased (Figure 3(b)). Continuing further in binding energy reveals several hexagonal/circular pockets surrounding the $\overline{\Gamma}$ point (Figure 3(c)). In order to better resolve the number of bands and their respective dispersion, we have plotted the dispersion cuts taken along the high-symmetry directions in the surface BZ (Figure 3(d)). Upon comparison with the CECs presented in Figure 3(a-c), we find two resolvable bands that extend to near the Fermi energy that make up the small pockets in Figure 3(a). Another parabolic hole-like band is seen with a band maximum just above $E-E_F\approx-0.6~\text{eV}$. These three bands are rather steeply dispersing around the $\overline{\Gamma}$ point. The $\overline{\Gamma}\overline{\text{M}}$ and the $\overline{\Gamma}\overline{\text{K}}$ on the left and right side of Figure 3(d) shows rather similar band dispersion near E$_F$. Below $E-E_F$ $\approx$ -0.8 eV the bands become less dispersive as they approach the $\overline{\text{M}}$ point, which produces the points of the hexagonal pocket in Figure 3(c). 

\indent The influence of the magnetic ordering on the electronic band structure was investigated by varying the temperature across T$_\text{N}$. Figure 3(e,f) shows the obtained ARPES spectrum along the $\overline{\text{M}}-\overline{\Gamma}-\overline{\text{M}}$ direction taken within the AFM phase (13.5 K) and the PM phase (30.0 K), respectively. In these panels, we normalized the momentum distribution curve intensities to better track the band dispersions. The overall band energies do not show strong modifications (i.e., shifting upwards or downwards) due to the onset of magnetic ordering; however, there are several small changes observed and highlighted by the red, blue, and black boxes. The black boxes highlight the apparent removal of a kink-like feature in the ARPES intensity, highlighting an increase in the band velocities extending up to the Fermi level in the AFM phase. Further down in binding energy, the blue and red boxes highlight two bands that appear to be pushed outward in momentum away from the $\overline{\Gamma}$ point. For further results on the temperature dependence, please refer to the supplementary materials \cite{SOM}. Although these modifications in the electronic band dispersion are rather small, their presence highlights coupling between the magnetic and valence electronic degrees of freedom. It is important to note that changes in the electronic structure across magnetic transition temperatures have also been observed in other magnetic topological quantum materials through temperature-dependent ARPES measurements \cite{Regmi_EuIn2As2, Sakhya_NdSb}

A comparison with calculated orbital-projected DFT band dispersions (Figure 3(g,h)) conducted within the antiferromagnetic phase, with Eu moments oriented along the $a$-axis as indicated in Figure 1(a), reveals a strong resemblance to our ARPES results. Figure 3(g) presents the calculated band dispersion and orbital projections onto the As $p$ orbital basis. From this, we find a significant As $p$ character to the dispersive valence bands. These calculations were performed by assuming a $U=5$ eV on-site interaction on the Eu 4$f$ sites. Treating the 4$f$ states as valence states, we find the Eu 4$f$ residing at around $E-E_F\approx$-1.5 eV binding energy, as demonstrated by the orbital projection onto the Eu $f$ orbitals in Figure 3(h). Consequently, a broad intensity is observed in the ARPES spectrum below about $E-E_F\approx-1~\text{eV}$, which is attributed to the Eu 4$f$ band intensity. The hybridization of the dispersive valence states with the Eu 4$f$ state is demonstrated to extend slightly beyond this intense flat band region.

The rather limited modifications of the dispersive bands near $E_F$ across the PM-AFM transition, presented in Figure 3(e,f), contrasts with reports in EuCd$_2$P$_2$ \cite{ZhangMIT}, where a full metal-insulator transition is observed. This limited response suggests that the resistivity anomaly in EuZn$_2$As$_2$ is driven by scattering of conduction electrons, rather than a change in the underlying electronic states. We note that our ARPES measurements were performed under zero applied magnetic field, and thus this conclusion primarily address the origin of the zero-field resistive anomaly. Within our experimental resolution, we observe only a weak electronic structure reconstruction below (T$_{\mathrm{N}}$), with no discernible modification of the bands in the immediate vicinity of ($E_{\mathrm{F}}$). At the same time, the field-dependent magnetotransport is likely governed by a more complex interplay of mechanisms. The application of magnetic field can modify the electronic eigenvalues through Zeeman splitting while simultaneously altering the magnetic structure, including the evolution from the canted antiferromagnetic to the field-polarized ferromagnetic state. Consequently, the magnetoresistance may reflect the combined influence of spin scattering, field-induced modifications of the electronic structure, and changes in the magnetic order. Elucidating the relative contributions of these effects will require future measurements capable of directly probing the electronic structure under applied magnetic field. Concerning the exchange mechanism coupling the valence electrons with the magnetic ordering, while magnetic interactions of lanthanide-based metals is usually described as having RKKY-type exchange coupling, it is interesting to note that, while our sample shows metallic behavior and a very small Fermi surface, many reports on other members of this family present a similar resistive anomalies in insulating samples within the Eu$X_2Pn_2$ family \cite{SinghSuperexchange,ZhangMIT,EuZn2P2FMMetal}. Scattering of Fermi level electrons from fluctuations of an RKKY-type exchange should show significant sensitivity on the Fermi surface size, and should disappear for insulating systems \cite{RKKY,NagaevLowCarrier}. Recently, an extended superexchange model has been invoked to describe the exchange coupling in insulating EuZn$_2$P$_2$ \cite{SinghSuperexchange}, which may also provide the mechanism for magnetic scattering in metallic members of this family as well \cite{NagaevLowCarrier}.\\
\indent Here we have investigated the overlap between the magnetic, transport, and electronic structure behavior in EuZn$_2$As$_2$. High-quality single crystals were successfully grown through Zn-As flux growth. Our magnetic characterization of the samples shows signatures of a PM-AFM phase transition at T$_\text{N}$ = 19 K, in agreement with previous results \cite{Blawat,BukowskiAFMOrder}. A significant resistive anomaly is observed for temperatures close to T$_\text{N}$, which is attributed to increased spin-disorder scattering of the carriers \cite{YamadaAFMMagnetoresistance, UsamiAFMMR}. This resistivity increase at the phase transition can be suppressed through the application of magnetic fields, both oriented within and perpendicular to the basal plane. In order to disentangle the scattering from the possibility of Fermi surface reconstruction that might contribute to the magnetotransport properties, we investigated the electronic structure both in the AFM and PM phases by utilizing high-resolution ARPES measurements combined with first-principles DFT-based calculations. While some modifications to the electronic bands are found, the extent of these changes are limited near the Fermi energy.\\

\section{Acknowledgements}

M.N. acknowledges support from the Air Force Office of Scientific Research MURI Grant No. FA9550-20-10322 and the National Science Foundation, Division of Materials Research, under Award No. 2518800. The work at Northeastern University was supported by the Air Force Office of Scientific Research MURI Grant No. FA9550-20-10322 and benefited from the computational resources of Northeastern University’s Advanced Scientific Computation Center (ASCC) and the Discovery Cluster. B.G. and K.G. acknowledge the National Supercomputing Mission (NSM) for providing computing resources of `PARAM RUDRA' at SNBNCBS, Salt Lake, Kolkata-700106, India. D.K. and T.R. were supported by the National Science Centre (Poland) under research grant 2021/41/B/ST3/01141. The authors acknowledge insightful discussions with Iftakhar Bin Elius and Nathan Valadez during this project. This work utilized resources of the Advanced Light Source at
the Lawrence Berkeley National Laboratory, a US Department of Energy Office of Science User Facility, under Contract No. DE-AC02-05CH11231. We thank Jonathan Denlinger for beamline assistance at the ALS Beamline 4.0.3.

\pagebreak

\title{Supplemental Material: Temperature Dependent Evolution of the Electronic Structure in EuZn$_2$As$_2$ across the N\'eel Transition}%

\maketitle

\section{Comparison of Magnetic AFMx and Nonmagnetic Band Structure Calculations}

ARPES spectra of Figures 3(e,f) from the main text are recreated here in Figure S1. Overlaid in blue lines are the DFT-calculated band eigenvalues (a) including and (b) excluding the $4f$ electrons, respectively. The localization in Eu 4$f$ electrons is critical to the magnetic moment formation. Calculations including the Eu 4$f$ orbitals produce a type-A AFM ordering. For modeling the non-magnetic reference state, a pseudopotential was used in which the Eu $4f$ states are treated as core states rather than valence states. Such an approximation is sometimes used as a practical way to obtain an approximate paramagnetic reference electronic structure for Eu-based compounds. Some of the changes to the ARPES spectra discussed in the main text are captured by including/excluding the $f$ electrons. For example, the development of a kink feature around $E-E_F=-0.3$ eV; however, any modifications tied to the PM-AFM transition will not be well captured below $E-E_F\approx$ -0.5 eV, where the $f$ weight is expected to become substantial.

\setcounter{figure}{0}
\renewcommand{\thefigure}{{S\textbf{\arabic{figure}}}}

\begin{figure}
    \centering
    \includegraphics[width=0.75\linewidth]{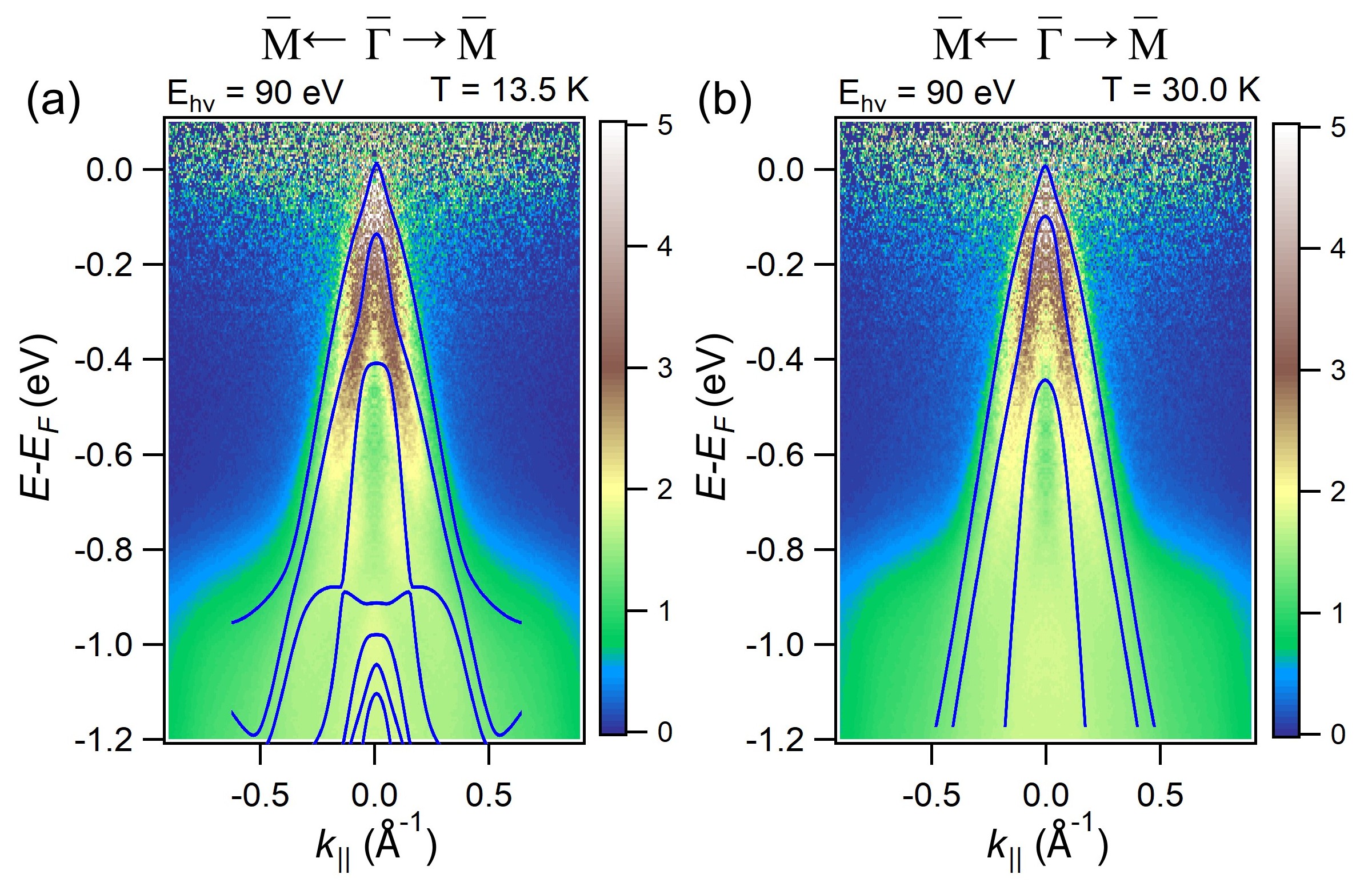}
    \caption{Effect of the Eu 4$f$ electrons. ARPES spectra of Figures 3(e,f) from the main text are recreated with DFT calculations (a) including and (b) excluding 4$f$ electrons, producing an AFM and NM ground state, respectively.}
    \label{fig:placeholder}
\end{figure}

\section{Eu 4$f$ Orbital Occupancy}

Figure. S2 explores the DFT-calculated Eu 4$f$ character. Panel S2(a) presents a diagram of the AFMx ground state obtained from self-consistent field calculations. Eu moments (pink) lie along the $a$-axis and orders in a type-A AFM configuration, with ferromagnetic couplings within the plane and antiferromagnetic couplings along the $c$-axis. Panels S2(b,c) present the DFT electronic band structure calculations, with the band eigenvalues presented with black lines and the Eu 4$f$ orbital contribution to the eigenstate indicated by the thickness of the teal highlight. An on-site interaction of U = 5 eV pushes the occupied 4$f$ bands to energies between $E-E_F$ = -1 eV to -1.8 eV, in close agreement with the core-level photoemission spectra presented in Figure 1(c) of the main text. The unoccupied 4$f$ orbitals are pushed approximately 8 eV above the Fermi energy, consistent with the well-defined Eu$^{2+}$ valency indicated by our Curie-Weiss fitting. The 4$f$ shell is half-occupied, resulting in 5 bands residing below $E_F$ and the remaining 5 bands residing above.

\begin{figure}
    \centering
    \includegraphics[width=1\linewidth]{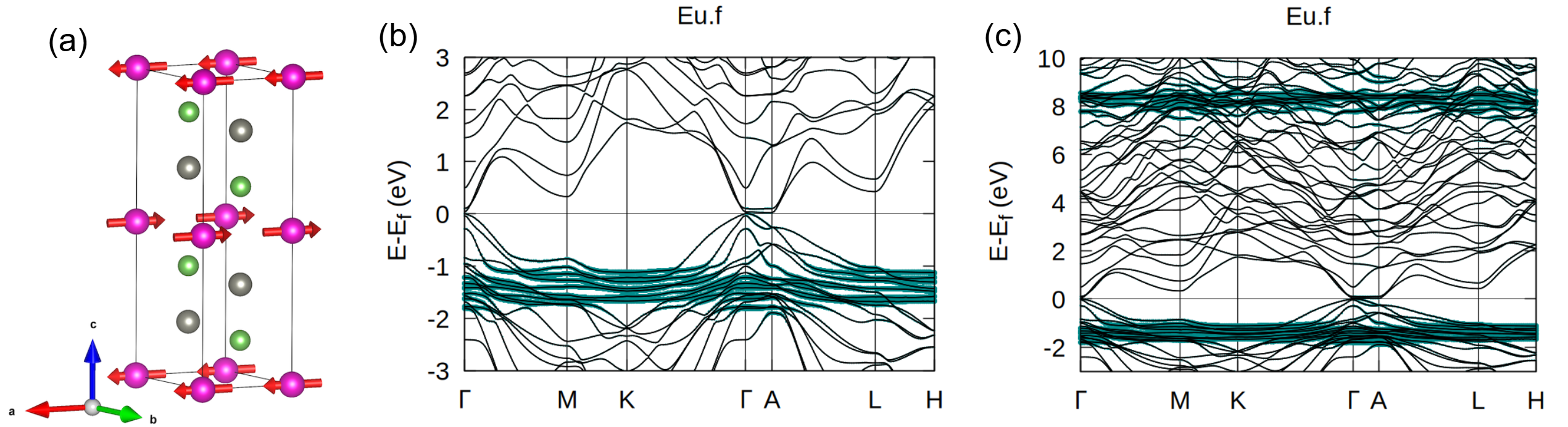}
    \caption{DFT-Calculated Eu 4$f$ Character. (a) Diagram of the AFMx ground state obtained from self-consistent field calculations. The Eu moments are indicated by pink arrows. (b,c) DFT electronic band structure calculations with the band eigenvalues presented with black lines and the Eu 4$f$ orbital contribution to the eigenstate indicated by the thickness of the teal highlight. (b) Presents the band structure over an energy window shown in the main text ($-3~\text{eV}<E-E_F<3~\text{eV}$), while (c) presents the band structure/$f$ orbital projection over a wider energy range from $E-E_F=-3$ eV to +10 eV.}
    \label{fig:placeholder}
\end{figure}

\section{Photon Energy Variation of ARPES along $\overline{\text{M}}-\overline{\Gamma}-\overline{\text{M}}$}

Figure S3 presents a comparison of the photoemission spectra across more photon energies and compares them between PM-phase measurements, taken at a temperature of 30 K, and AFM-phase measurements, taken at a temperature of 13 K. First, the $k_z$ dispersion is evident by comparing the spectra across photon energies, particularly in whether the valence band maximum crosses the Fermi energy. Across different photon energies, we see no evidence of in-plane AFM backfolding, consistent with a $k_z$-folding induced by the out-of-plane magnetic propagation vector.

\begin{figure}
    \centering
    \includegraphics[width=1\linewidth]{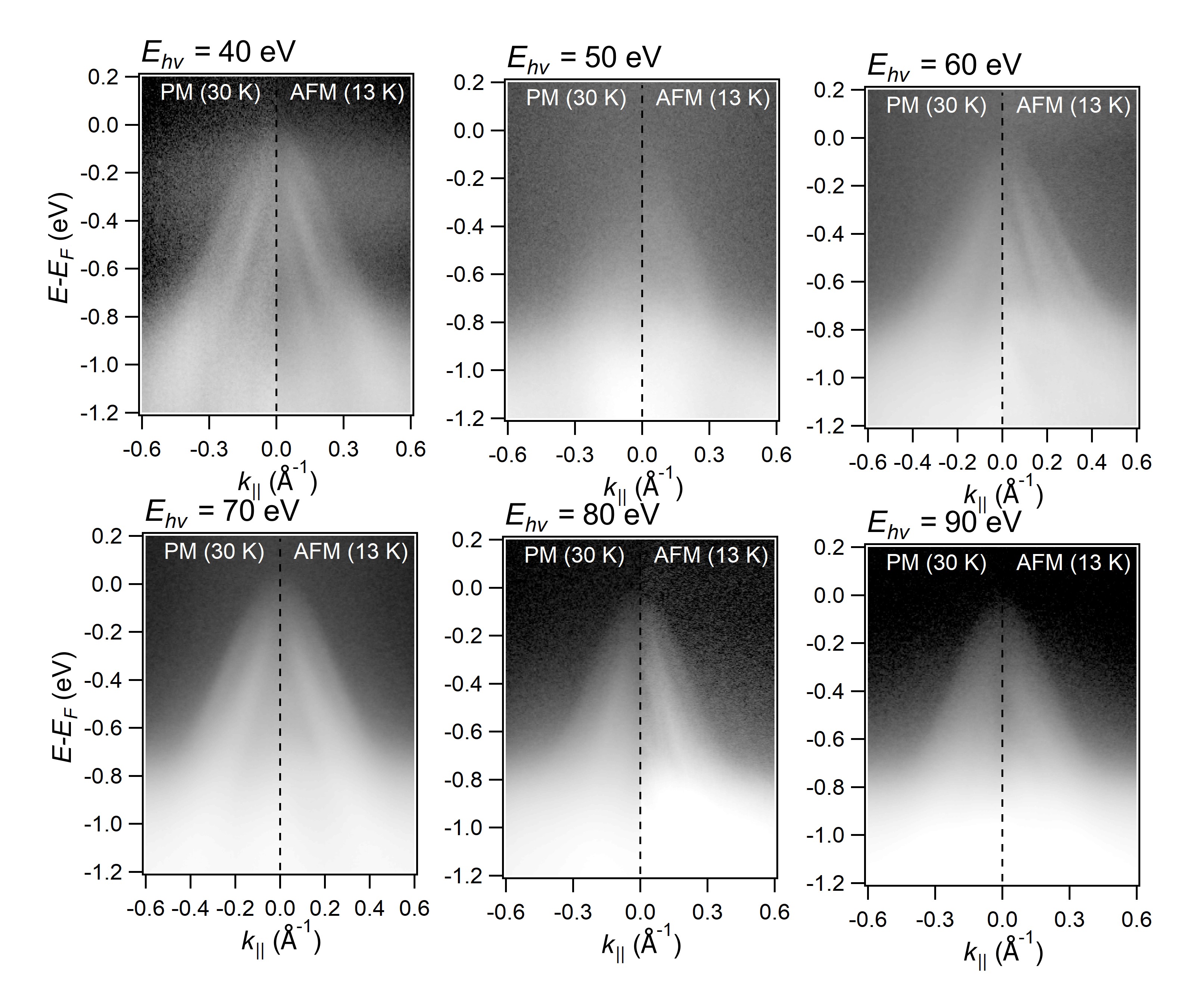}
    \caption{Photon energy dependence of the ARPES-obtained band dispersions along $\overline{\text{M}}-\overline{\Gamma}-\overline{\text{M}}$ direction. The ARPES spectra presented here were acquired at photon energies ranging from 40 eV to 90 eV in 10 eV increments. In each panel, the photon energy is indicated by $E_{h\nu}$ on the top left, and the spectra is split into results from the PM phase (T = 30 K) on the left half, and AFM phase (13 K) on the right half.}
    \label{fig:placeholder}
\end{figure}

	\begin{figure*}[t]
	\centering
	\includegraphics[width=1\linewidth]{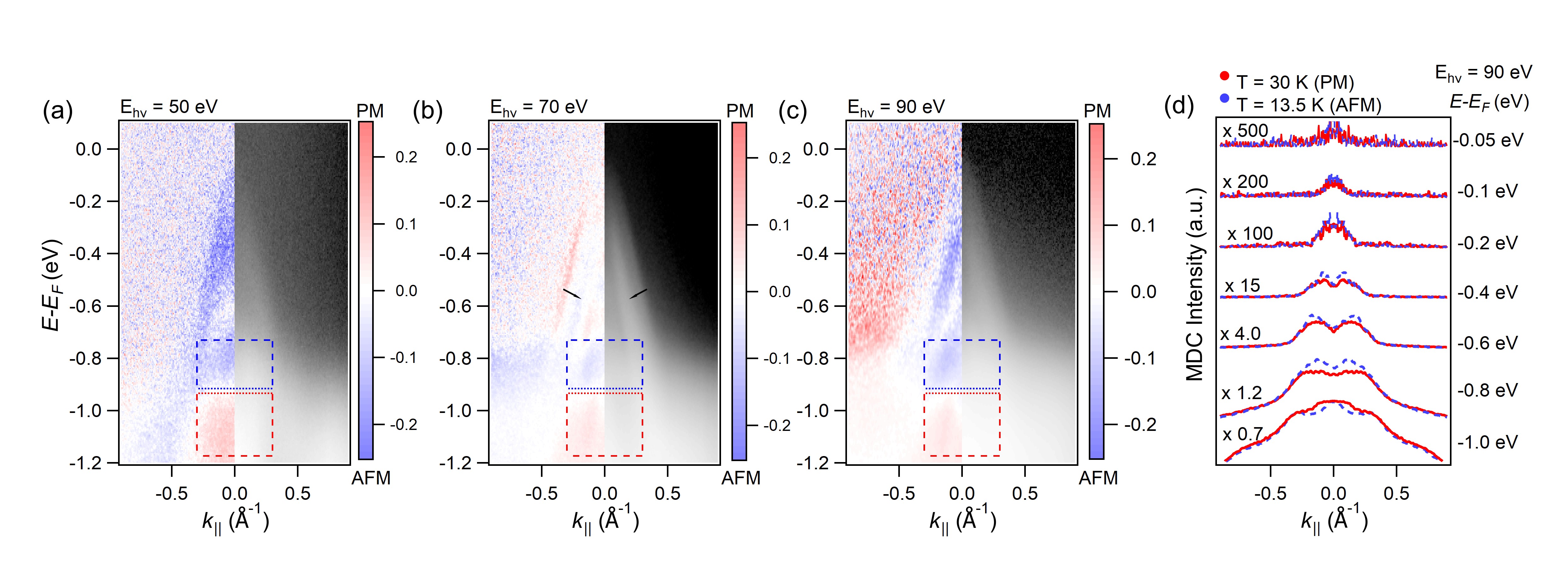}
	\caption{Evolution of the electronic structure across T$_\text{N}$. The left side of panels (a-c) present the normalized difference between AFM and PM photoemission intensities along the $\overline{\text{M}}-\overline{\Gamma}-\overline{\text{M}}$ direction at incident photon energies of (a) $E_{h\nu} = $ 50 eV, (b) 70 eV, and (c) 90 eV. The blue color indicates regions where the spectral weight is stronger in the AFM phase, while indicates where it is stronger in the PM phase. The right half of each panel presents the raw ARPES data measured at a temperature T = 13.5 K. (d) Photoemission intensity momentum distribution curves (MDCs). The PM-phase MDCs are shown in red and the AFM MDCs are given in blue. The MDCs have been rescaled by factors ranging from 0.7$\times$ for $E-E_F = -1.0~\text{eV}$ and 500$\times$ for $E-E_F = -0.05~\text{eV}$, as indicated on the left side of the panel.}
	\label{fig:layout4v2}
\end{figure*}

\section{Temperature-Dependent Photoemission Intensity Shifts}
Another way to visualize these changes is to directly plot the MDCs, as presented in Figure S4(d). The MDCs are taken along the same $\overline{\text{M}}-\overline{\Gamma}-\overline{\text{M}}$ direction, with red indicating the measurements from the PM phase (T = 30 K) and blue indicating those from the AFM phase (T = 13.5 K). To aid the visualization of the MDCs near the Fermi energy, we have modified the relative amplitudes by the factors shown on the left of panel S4(d), ranging from $\times0.7$ to $\times500$. While clear shifts in the MDC weights are observed, corresponding with the shift in band centers discussed before, the relative intensities of the MDCs seem to be enhanced correlating with the magnetic phase. Specifically, we see an enhancement of the mid-ranged features, between $E-E_F=$ -0.2 eV -- -0.8 eV, within the AFM phase, where the MDC peaks appear to be sharper and more intense. For $E-E_F=$ -1.0 eV, we find the opposite trend, where PM photoemission intensity is stronger at this binding energy.\\
\indent To assess the extent these temperature-dependent modifications vary with photon energy, we present the ARPES dispersion cuts along $\overline{\text{M}}-\overline{\Gamma}-\overline{\text{M}}$ measured using $E_{h\nu}$ = 50 eV, 70 eV, and 90 eV in Figure S4(a-c), respectively. Variation of the incident photon energy results in the dispersion of the surface-projected electronic bands along $k_z$ perpendicular to the cleaved surface. Due to the sizeable $k_z$-broadening, the ARPES-obtained dispersion does not show much variation with photon energy. However, taking the difference between PM-phase and AFM-phase spectra reveals somewhat varying temperature-dependent modifications of the electronic spectral function. The left half of the panels in Figures S4(a-c) presents the normalized-difference, $$\Delta I(k,E) = \frac{I_{30~\text{K}}-I_{13.5~\text{K}}}{I_{30~\text{K}}+I_{13.5~\text{K}}},$$ between the spectra measured at sample temperatures of 30 K (PM phase), represented by the photoemission intensity $I_{30~\text{K}}$, and 13.5 K (AFM phase), with intensity $I_{13.5~\text{K}}$, taken using 90 eV photon energy. The red coloring indicates a stronger photoemission intensity ($\Delta I >0$) in the PM phase, while blue indicates a stronger intensity in the AFM phase ($\Delta I < 0$)). The right half of each panel shows the raw photoemission spectrum. At $E_{h\nu}$ = 50 eV (Figure S4(a)), the bands between $E-E_F$ = 0.9 eV and $E_F$ show increased photoemission intensity within the AFM phase, while at $E_{h\nu}$ = 70 eV (Figure S4(b)) there appears to be a shift of the bands toward smaller momentum, as indicated by the arrows pointing from the red (PM) toward the blue (AFM) region. At $E_{h\nu}$ = 90 eV there again seems to be an increase in the intensity of these bands in the AFM phase, however the solid white regions indicate a strong overlap in the observed photoemission between the PM and AFM spectra for the outermost bands. In contrast to the change in behavior with photon energy for the bands above $E-E_F>$ -0.7 eV, we find a more consistent shift of the spectral weight from within the red boxes in Figure S4(a-c) in the PM phase, to the blue boxes in the AFM phase for $E-E_F<$ -0.7 eV. For $E-E_F<-1.0 ~eV$, the photoemission intensity for the T = 30 K (PM) measurement was higher than the T = 13 K (AFM) one, while for $-1.0~eV<E-E_F<-0.6~eV$ the case is reversed.\\
\indent This constant shift of the spectral weight upward from the red box to the blue box upon crossing into the AFM phase may be better understood by considering the orbital composition of these bands. We have performed orbital-projected DFT-based calculations, which attributes an As $p$ and Eu $f$ orbital composition to the observed electronic states, as seen in Figures 3(g) and (h) of the main text, respectively. Considering the orbital compositions, the upward shift of the spectral weight upon the onset of magnetic ordering might potentially be taken as an indication of a reduction in $pf$ hybridization as the magnetic degree of freedom locks into place.

\end{document}